# Reversible hydrogen storage capacity of Sc and Y functionalized [1,1]paracyclophane: Insights from density functional study

Rakesh K. Sahoo[1], P. Kour[2], Sridhar Sahu[1,*]

Email: sridharsahu@iitism.ac.in

[1]Computational Materials Research Lab, Department of Physics, Indian Institute of Technology (Indian School of Mines) Dhanbad, India

[2] Department of Physics, Birla Institute of Technology, Mesra, Patna Campus, Patna, 800014, India

## Abstract

This work reports the hydrogen storage, and delivery capacities of Sc and Y functionalized [1,1]paracyclophane using dispersion corrected density functional theory calculations. The Sc and Y atoms are bind strongly with benzene rings of [1,1]paracyclophane. Each Sc and Y atom functionalized over [1,1]paracyclophane adsorb up to 6 $H_2$ molecules via Kubas interaction achieving maximum gravimetric density up to 8.22 wt% and 6.33 wt%, respectively. The calculated average hydrogen adsorption energy (0.36 eV) is lower than the chemisorption but higher than the physisorption process. The kinetic stabilities are verified through the HOMO-LUMO gap and different global reactive descriptors. ADMP molecular dynamics simulations reveal the reversibility of adsorbed $H_2$ molecules at sufficiently above the room temperature and the solidity of host material at 500 K. Average Van't Hoff desorption temperature for Sc and Y decorated system was calculated to be 439 K and 412 K respectively at 1 atm of pressure. The estimated thermodynamically usable hydrogen capacities are found to be 5.92 wt% and 5.87 wt% of hydrogen which fulfil the hydrogen energy criteria (by 2025) of US-DOE. Hence, we believe that Sc and Y functionalized [1,1]paracyclophane can be considered as a thermodynamically viable, and potential reversible hydrogen storage materials at ambient environment.

**Keywords:** Hydrogen storage, DFT, Van't-Hoff equation, ADMP, [1,1]paracyclophane, ESP

## 1. INTRODUCTION

The excessive demand on fossil fuels for vehicular and industrial purposes has not only caused depletion of the limited fossil fuels but also triggered alarming global warming and environmental pollution [1]. Therefore, international organizations such as United Nation Environment Program (UNEP) has been urging for alternative and clean energy resources. In this regard, hydrogen is considered as an ideal, environment-friendly, and sustainable energy carrier, which, in addition, has higher energy per unit mass (120 MJ/kg) than the fossil fuels (44 MJ/kg for gasoline) [2,3] However, efficient storing and delivering hydrogen energy as fuel under ambient conditions, has been a major challenge for the commercial developments. [4,5]. To store hydrogen in compressed gaseous form, voluminous tanks are required that can withstand high pressure (~70 MPa), and which is quite difficult and expensive [6]. Storing liquid hydrogen requires the cryogenic temperature about -253°C under high pressure (~ 250-350 atm) which makes the process highly expensive [7]. Therefore, solid-state material-based hydrogen storage systems have been proposed to be the selective method which is supposed to efficient and cost-effective as well. The basic criteria that an efficient solid-state storage system should follow, according to the US department of energy (DOE-US) are; a) intermediate adsorption energy in the range of 0.2-0.6 eV/$H_2$ b) minimum gravimetric density up to 5.5 wt%, c) volumetric density above 40 kg/$m^3$ d) refuelling time of less than 3 minutes over the lifetime, and d) fast kinetic with ambient thermodynamics (T ~40-85 °C and under P ~100 atm) [8,9]

Since last few decades, researchers have extensively studied different substrates such as carbon nanostructures [10,11,12], metal hydrides [13,14], graphene [15,16], zeolites [17], metal-organic frameworks [18,19], as hydrogen storage materials. However, these storage systems are reported have several drawbacks such as low storage capacity, high desorption temperature, instability at high temperature. So, these issues must be addressed while developing viable hydrogen storage substrates for commercial applications. Organometallic compounds functionalized with transition metals (TM) such as TM-decorated organometallic buckyballs [20], TM-ethylene [21], can be highly potential hydrogen storage substrates as compared to many other materials reported in literature. It was reported that TM bonded strongly with π -electron delocalized compound via Dewar coordination and trap hydrogen molecules through Kubas interaction [22, 23]. Many authors have reported TM atoms like Ti, Sc and Y functionalized carbon-based nanostructures for efficient hydrogen storage [24, 25, 26, 27, 28, 29] For, example Modak et al. made a comparative study of the hydrogen uptake capacity of single-wall carbon nanotubes functionalized with Y, Zr, Nb, and Mo. Their study suggested that, TM having lowest number of d electron in their outer shell showed better hydrogen storage capacity [30]. Mananghaya *et al.* studied hydrogen storage in Sc and Ti coating single walled carbon nanotubes using thermodynamic simulations and reported a gravimetric capacity of 5.85 wt% at 300K [31]. Recently, the hydrogen storage capacity of scandium and yttrium decorated $C_{24}$ fullerene was studied by Shukla *et al*. [32, 33]. They found that, each Sc and Y atom could capture six hydrogen molecules leading to maximum gravimetric density up to 13.02 wt% at high temperature of 500 K. Many researchers have studied the hydrogen storage properties and capacities of MOF impregnated with fullerenes [34,35]. Foe example, Rao *et al.* investigated $H_2$ storage capacity of Li doped MOF impregnated with Li-coated fullerenes and reported a gravimetric capacity of 6.3 wt% at 100 bar and 243 K [36]. Yu *et al.* studied the hydrogen gravimetric and volumetric capacity of fullerene impregnated with IRMOFs and reported $C_{60}$ in IRMOF had a capacity of 7.4 wt% at 77K and 18 bar [37]. Mehrabi et al. experimentally showed that Pd doped multi-walled carbon nanotubes are potential hydrogen storage system with nearly 6wt% hydrogen capacity [38]. Chakraborty *et al.* investigated the $H_2$ storage capabilities of yttrium coated carbon nanotube and reported 6.1 Wt% of storage capacity with 100 % desorption of $H_2$ at 612 K [39]. Zhang *et al.* studied yttrium doped $B_{40}$ and reported that each Y atom could adsorb $5H_2$ molecules leading to 5.8 wt% with an adsorption energy of -0.211 eV/$H_2$ [40]. Sathe et al. studied the hydrogen storage in Li and Sc functionalized [4,4]paracyclophane and reported that each Sc atom can adsorb via physisorption of $5H_2$ leading to 11.8 wt% of storage capacity [41]. They showed that Sc atoms are bound with PCP44 via Dewar mechanism, and the hydrogen molecules are attached to the sorption canter via Kubas interaction. Kumar et al. reported the hydrogen storage capacity up to 10.3 wt% of paracyclophane decorated with Sc and Li [42].

In this report, we have explored the hydrogen uptake and delivery capacity of [1,1]paracyclophane (PCP11) functionalized with Sc and Y transition metal atoms. The preceding square bracket number, "[1,1]" in [1,1]paracyclophane, indicates that the consecutive benzene rings (2 benzene rings) in paracyclophane are linked with one (-$CH_2$-) moiety [43]. The linking bridges are relatively short; thus, the separation between consecutive benzene rings is small which develops a strain in the aromatic rings. This strain in the rings can be utilized for Sc and Y functionalization over the aromatic benzene ring. Due to the strain and metal functionalization, the aromatic benzene rings losses their inherent planarity [44]. Since, aromatic benzene rings are there in the PCP, these are easy for experimental synthesis and functionalization of metal atoms [45]. From the previous reports it is evident that metal functionalized with aromatic benzene rings could be treated as a viable hydrogen storage material [46, 47]. Recently, Sathe et al. investigated the hydrogen adsorption properties of Li functionalized [1,1]paracyclophane, and reported that each Li atom on PCP11 could hold $4H_2$ molecules with a gravimetric density of 13.42 wt% [48]. We choose to functionalize Sc and Y transition metal atoms over the PCP11, as both are TM atom possess minimum number of electrons in d- block which can be beneficial for the reversible hydrogen adsorption. Our estimation reveals that each Sc and Y atoms on PCP11 can adsorb $6H_2$ molecules via Kubas interaction. We verified the solidity of host materials and theoretically predicted hydrogen storage capacities of these materials by using molecular dynamics simulations. We believe that our work on Sc and Y functionalized [1,1]paracyclophane as potential hydrogen storage material can contribute substantially to the research and development of new materials for hydrogen energy.

## 2. THEORY AND COMPUTATION

The important mathematical parameters to quantify the characteristics of the hydrogen storage systems are the average binding energy of the host clusters, adsorption and successive desorption of hydrogen molecules from the host substrates.

The binding strength of the transition metal atom (Sc and Y) on the PCP11 is calculated using the following equation.

$$E_b = \frac{1}{m}[E_{PCP11} + mE_{TM} - E_{mTM+PCP11}] \qquad (1)$$

Where $E_{PCP11}$, $E_{TM}$, and $E_{mTM+PCP11}$ are the total energy of PCP11, metal atom and TM-decorated PCP11 respectively. m is the number of metal atoms added over PCP11 substrate.

Then the average adsorption energy per hydrogen molecule of the system can be calculated as;

$$E_{ads} = \frac{1}{n}\left[E_{PCP11+TM} + nE_{H_2} - E_{mTM+PCP11+nH_2}\right] \quad (2)$$

Where $E_{PCP11+TM}$, $E_{H_2}$, and $E_{mTM+PCP11+nH_2}$ is the total energy of host material, hydrogen molecule and hydrogen trapped complexes respectively. n is the number of $H_2$ molecules adsorbed over each complex.

Now, at a particular thermodynamic condition, the successive desorption energy of adsorbed $H_2$ molecules is calculated using the following equation.

$$E_{des} = \frac{1}{n}\left[2E_{H_2} + E_{Host+(n-2)H_2} - E_{Host+nH_2}\right] \quad (3)$$

Here, (n -2) factor in the above formula arises because of successive desorption of one $H_2$ molecule each from the two TM adsorbents. The above energetic calculations have been performed incorporating the basis set superposition error (BSSE).

Moreover, computations of gravimetric density and thermodynamics of the studied storage systems are also important in view of their practical usability.

To obtained the hydrogen uptake capacity, gravimetric density (wt%) of hydrogen can be calculated using the following equation:

$$H_2(wt\%) = \frac{M_{H_2}}{M_{H_2}+M_{Host}} \times 100 \quad (4)$$

Here $M_{H_2}$ represent the mass of the total number of $H_2$ molecules adsorbed and M Host represent the mass of metal-doped PCP11.

To depict a quantitative picture of $H_2$ adsorption and desorption at different temperature (T) and pressure (P), the number of $H_2$ molecules adsorbed on each sorption center (occupation number) "N" in TM decorated PCP11 is calculated by [49, 50].

$$N = \frac{\sum_{n=0}^{n_{max}} n g_n e^{[n(\mu-E_{ads})/K_BT]}}{\sum_{n=0}^{n_{max}} g_n e^{[n(\mu-E_{ads})/K_BT]}} \quad (5)$$

Here $n_{max}$ is the maximum number of $H_2$ molecules adsorbed in each TM on PCP11, n and g n represent the number of $H_2$ molecules adsorbed and configurational degeneracy (taken $g_n$ = 1 by avoiding phonon contribution to entropy) for a n. $K_B$ is the Boltzmann constant and E ads (<0) indicates the adsorption energy of $H_2$ molecules to TM-PCP11. µ is the chemical potential of $H_2$ at specific T and P, obtained by using the following expression [51].

$$\mu = H^0(T) - H^0(0) - TS^0(T) + K_BT \ln\left(\frac{P}{P_0}\right) \quad (6)$$

Here $H^0(T)$, $S^0(T)$ are the enthalpy and entropy of $H_2$ at pressure $P^0$(1 bar).

The bare and hydrogenated TM-decorated [1,1]paracyclophane (PCP11) complexes are optimized using hybrid ωB97Xd functional along with 6-311+G(d,p) basis sets within the framework of density functional theory. ωB97Xd is a range separated version of Becke's 97 functional and includes the long-range and Grimme's D2 dispersion correction [52, 53]. It is worth mentioning that ω B97Xd approach is a reliable method for investigating non-covalent interaction, organometallic systems, and its thermochemistry. During the geometry optimization of metal functionalized PCP11, all electron triple-split valence basis set 6-311+G(d,p) is used for carbon, hydrogen and scandium atom, whereas, for the yttrium atom, an effective core potential (ECP) based LanL2DZ basis set is used [54]. The harmonic frequencies of all the studied systems are calculated to ensure the systems are in true ground state on the potential surface.

The atomistic molecular dynamic (AMD) simulations are then performed employing the extended Lagrangian technique, atom-centered density matrix propagation (ADMP), to explore the solidity of host materials and the reversibility of hydrogen molecules from the substrate. The time step for ADMP molecular dynamics simulations is set at 1 fs, and using the velocity scaling method, the temperature is maintained throughout the simulations. All the calculations are performed with gaussian 09 computational program [55].

## 3. RESULTS AND DISCUSSION

Optimized structure of PCP11 is shown in Figure 1(a). There are two benzene rings in PCP11, connecting with single -$CH_2$- moiety as bridge. The distance between carbon atoms of -$CH_2$- and benzene is 1.55 Å, which corresponds to the findings of Sathe et al. [48]. The attached benzene rings lose their inherent planarity and bend slightly inwards due to the short connecting bridges and the strain induced by it. The induced strain also has a vital role in transition metal (TM) functionalization. Before functionalizing the TM atom, we have calculated the Nucleus Independent Chemical Shift (NICS) in order to check the aromaticity of the benzene rings. NICS value is calculated from the center to 3 Å above benzene ring with an increment of 1 Å. At a distance of 1 Å above the benzene ring, NICS is found negative maximum (–8.66 ppm), indicating the high aromaticity [56, 57]. The optimization of TM-functionalized PCP11 is then initialized for NICS(1) of the benzene rings.

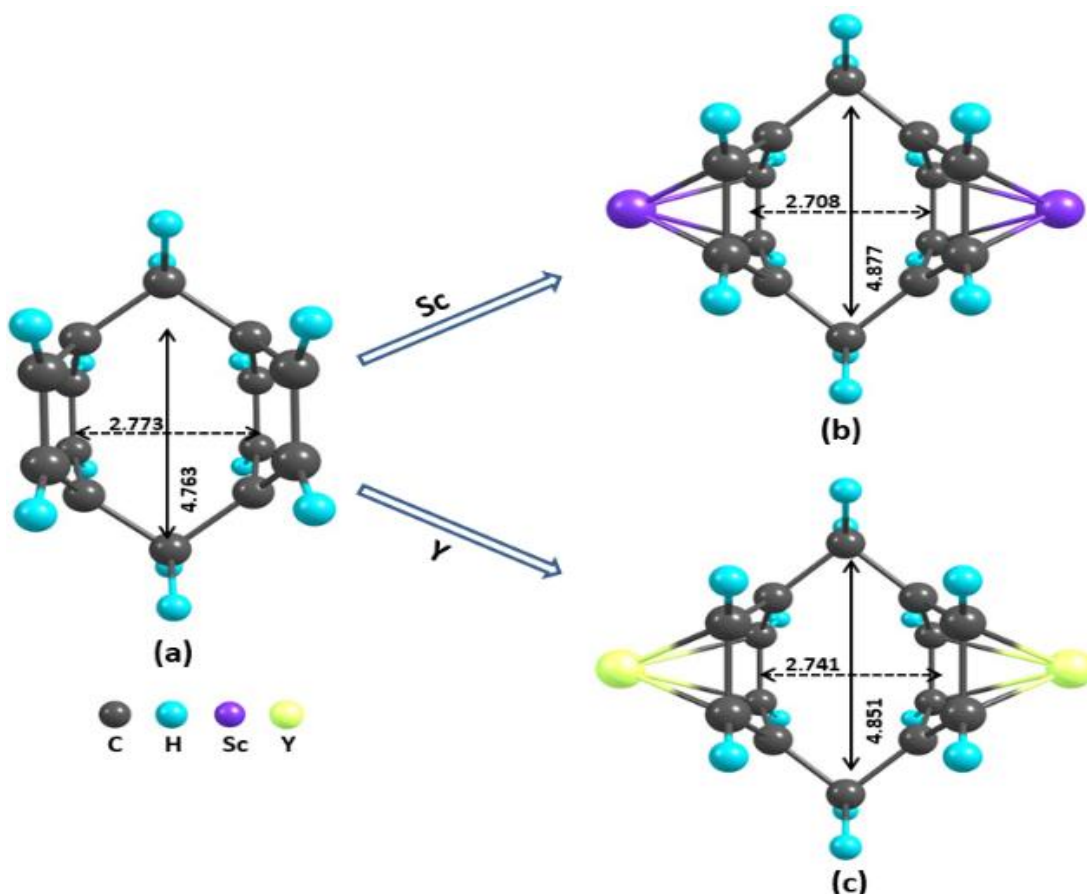

**Figure 1:** Optimized structures of (a) PCP11, (b) PCP11-2Sc, and (c) PCP11-2Y

### 3.1. Functionalization of Sc and Y

Optimized structures of Sc and Y functionalized over PCP11 (PCP11-2Sc, and PCP11-2Y) are shown in Figure 1(a) and 1(b) respectively. The Sc and Y atoms are grafted over the benzene ring of PCP11 with average binding energy of 1.43 eV and 1.66 eV, respectively. These high binding energy value reveals the strong bonding of Sc and Y with PCP11 and averts the possibility of metal clustering. The Sc and Y atoms are supposed to bind with the delocalised $\pi$-electrons of the benzene ring of PCP11 via Dewar coordination. In Dewar coordination, the benzene ring of the system donates its $\pi$-electrons density from the bonding orbital to empty d-orbital of Sc and Y atoms. Further, the electrons from filled d-orbital are back donated to empty $\pi^*$-antibonding orbital of the benzene ring of PCP11. The average distance of Sc from the center of benzene ring ($R_c$) is found as 1.814 Å and the C-C bridge distance reduced to 1.52 Å after metal functionalization, which matches with earlier reports [41, 59]. Similarly, the average distance between Y and $R_c$ is found to be 2.047 Å and C-C bridge bond length reduced to 1.525 Å. The highest occupied molecular orbital (HOMO) and lowest unoccupied molecular orbital (LUMO) energy gap ($E_g$) of PCP11 is found to be 7.26 eV. This energy gap is lowered to 3.02 eV and 2.45 eV after functionalization of Sc and Y atoms, respectively. Reduction in the energy gap of the substrate may be due to the contribution of d orbital of TM atoms [58, 59]. The high energy gap of these systems reveals their kinetic stabilities. The electrostatic potential (ESP) map of bare and TM functionalized PCP11 has been plotted and depicted in Figure S1 (*in supporting information*). In ESP plot, the red region over benzene ring shows the accumulation of electron density and the blue region over Sc and Y region indicates the deficiency of electron density. The calculated average Hirshfeld charge on the benzene ring of PCP11 is -0.034 e.u. The Hirshfeld charge on Sc and Y atom after functionalization is +0.423 e.u and +0.399 e.u, respectively, which makes TM atoms ionic. These ionic open metal sites can adsorb the guest $H_2$ molecules.

**Table 1:** Average distance between carbon bridge (C-C), center of PCP11 benzene ring ($R_c$) and Sc atom ($R_c$-Sc), Sc and adsorbed hydrogen molecules (Sc-H), and the intermolecular hydrogen distances (H-H).

| Name of complex | Bridge c-c (Å) | $R_c$-Sc (Å) | Sc-H (Å) | H-H (Å) |
|---|---|---|---|---|
| PCP11-2Sc | 1.520 | 1.814 | | |
| PCP11-2Sc-2$H_2$ | 1.522 | 1.855 | 1.982 | 0.807 |
| PCP11-2Sc-4$H_2$ | 1.527 | 1.937 | 1.963 | 0.821 |
| PCP11-2Sc-6$H_2$ | 1.532 | 1.966 | 2.091 | 0.805 |
| PCP11-2Sc-8$H_2$ | 1.526 | 2.039 | 2.004 | 0.806 |
| PCP11-2Sc-10$H_2$ | 1.525 | 2.055 | 2.020 | 0.795 |
| PCP11-2Sc-12$H_2$ | 1.525 | 2.050 | 2.473 | 0.786 |

**Table 2:** Average distance between carbon bridge (C-C), center of PCP11 benzene ring ($R_c$) and Y atom ($R_c$-Y), Y and adsorbed hydrogen molecules (Y-H), and the intermolecular hydrogen distances (H-H)

| Name of complex | Bridge c-c (Å) | $R_c$-Y (Å) | Y-$H_2$ (Å) | H-H (Å) |
|---|---|---|---|---|
| PCP11-2Y | 1.525 | 2.047 | | |
| PCP11-2Y-2$H_2$ | 1.523 | 2.067 | 2.178 | 0.796 |
| PCP11-2Y-4$H_2$ | 1.528 | 2.143 | 2.136 | 0.818 |
| PCP11-2Y-6$H_2$ | 1.532 | 2.180 | 2.184 | 0.804 |
| PCP11-2Y-8$H_2$ | 1.528 | 2.226 | 2.194 | 0.799 |
| PCP11-2Y-10$H_2$ | 1.527 | 2.236 | 2.284 | 0.789 |
| PCP11-2Y-12$H_2$ | 1.525 | 2.270 | 2.372 | 0.781 |

### 3.2. Geometry and stability after $H_2$ adsorption

After functionalization of Sc and Y atoms on PCP11, hydrogen molecules are introduced to each TM atom sequentially. Sequential addition of $H_2$ atoms to each sorption center which lowers the intricacy of steric hindrance due to $H_2$ crowed, urges to investigate the stability of each hydrogenated complex as well as the pattern of adsorption and desorption energy.

The successive addition of $H_2$ molecules in PCP11-2Sc reveals that each Sc atom can hold up to six $H_2$ molecules. The hydrogenated complexes PCP11-2Sc-2n$H_2$ (n= 1-6) are depicted in Figure 2. The average distance between Sc atom and $R_c$, Sc and adsorbed $H_2$ molecules (Sc-$H_2$), and inter-molecular hydrogen distance (H-H) are provided in Table 1. It is found that the Sc-$R_c$ and Sc-$H_2$ distances increases with the number of hydrogen molecule in the complexes. On introduction of first $H_2$ to PCP11-2Sc the Sc-$R_c$ bond distance increase by 2 %, while on addition of sixth hydrogen this distance increases by almost 13%. This increase in bond distance is due to small charge transfer from the benzene ring of PCP11 to Sc atoms. Further the average Sc-$H_2$ distances also increases from 1.98 Å to 2.47 Å after adsorption of sixth hydrogen to each Sc which may be attributed to the steric hindrance due to hydrogen crowed around the Sc atom. The H-H bond length (0.82 Å) is found to be elongated by 10% from experimental value of isolated $H_2$ (0.74 Å) which indicate hydrogen undergoes molecular adsorption on PCP11-2Sc. Similarly, Sequential adsorption of hydrogen on PCP11-2Y show that each Y atoms can bind up-to 6$H_2$ molecules, and the optimized structures of $H_2$ adsorbed systems PCP11-2Y-2n$H_2$ (n= 1-6,) are depicted in Figure 3. The geometrical bond length of Y-$R_c$ and Y-$H_2$ and H-H are presented in Table 2. The distance between adsorbed $H_2$ molecules and the sorption center increases with increase in hydrogen molecules in the system. The inter-atomic hydrogen distance is found in the range of 0.78 Å to 0.81 Å.

In order to investigate the stability of bare and hydrogenated systems, we have estimated various reactivity parameters such as the energy gap between Highest Occupied Molecular Orbital ((HOMO) and Lowest Unoccupied Molecular Orbital (LUMO) ($E_g$), hardness (η), electrophilicity index (ω), using the Koopman's theorem [60]. A larger $E_g$ corresponds to a higher energy required for an electron jump from HOMO to LUMO. In other words, a smaller $E_g$ gap suggests greater chemical reactivity, whereas a larger $E_g$ gap indicates decreased chemical reactivity. The variation of HOMO-LUMO gap with the number of hydrogen molecules in PCP11-2Sc(2Y) are depicted in Figure 4. It is seen that the $E_g$ shows an increasing tendency with increase in number of hydrogen molecules on the host, which reveals the kinetic stability of the studied systems with successive addition of $H_2$ molecules (Figure 4). It has been observed that, the value of η is decreases with increase in number of $H_2$ molecules in the system. Similarly, ω value decreases with the rise in value of $H_2$ in the system (Figure: S2 and Table: S1). This assures the stabilities of the studied systems by following the *maximum hardness and minimum electrophilicity principle* [61].

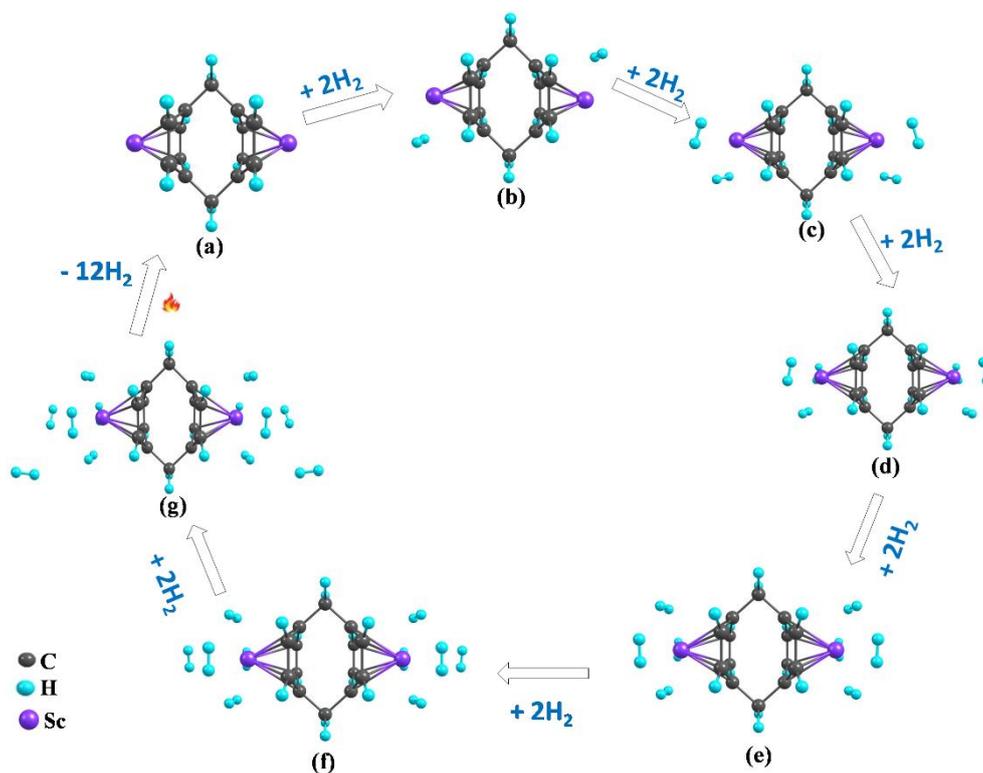

**Figure 2:** Optimized geometry of bare and hydrogenated Sc functionalized PCP11
(a) PCP11-2Sc, (b) PCP11-2Sc-2H$_2$, (c) PCP11-2Sc-4H$_2$, (d) PCP11-2Sc-6H$_2$, (e) PCP11-2Sc-8H$_2$, (f) PCP11-2Sc-10H$_2$, (g) PCP11-2Sc-12H$_2$

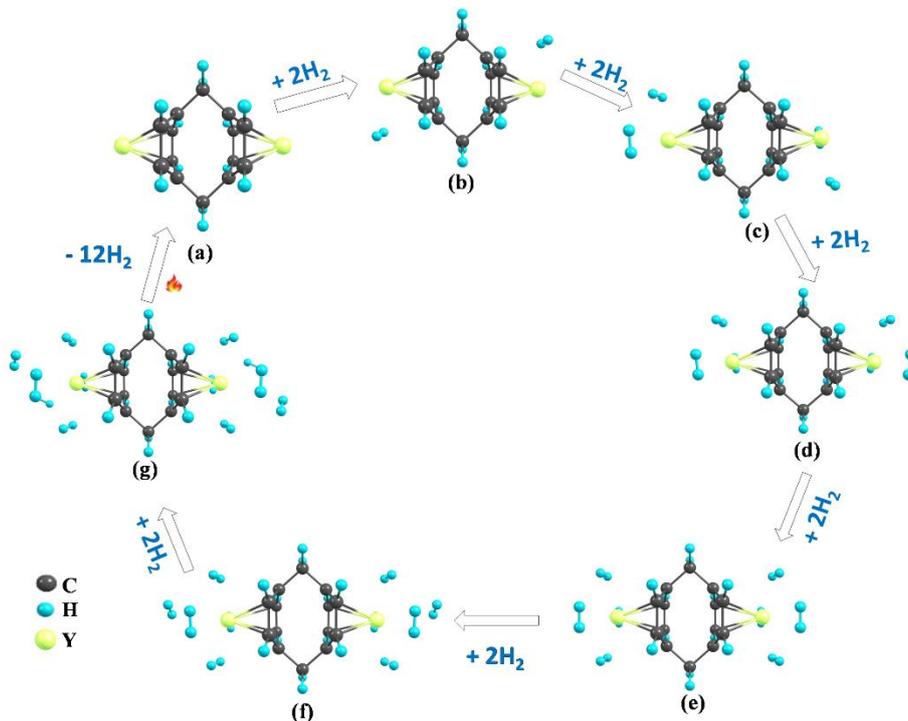

**Figure 3:** Optimized geometry of bare and hydrogenated Sc functionalized PCP11
(a) PCP11-2Y, (b) PCP11-2Y-2H$_2$, (c) PCP11-2Y-4H$_2$, (d) PCP11-2Y-6H$_2$, (e) PCP11-2Y-8H$_2$, (f) PCP11-2Y-10H$_2$, (g) PCP11-2Y-12H$_2$

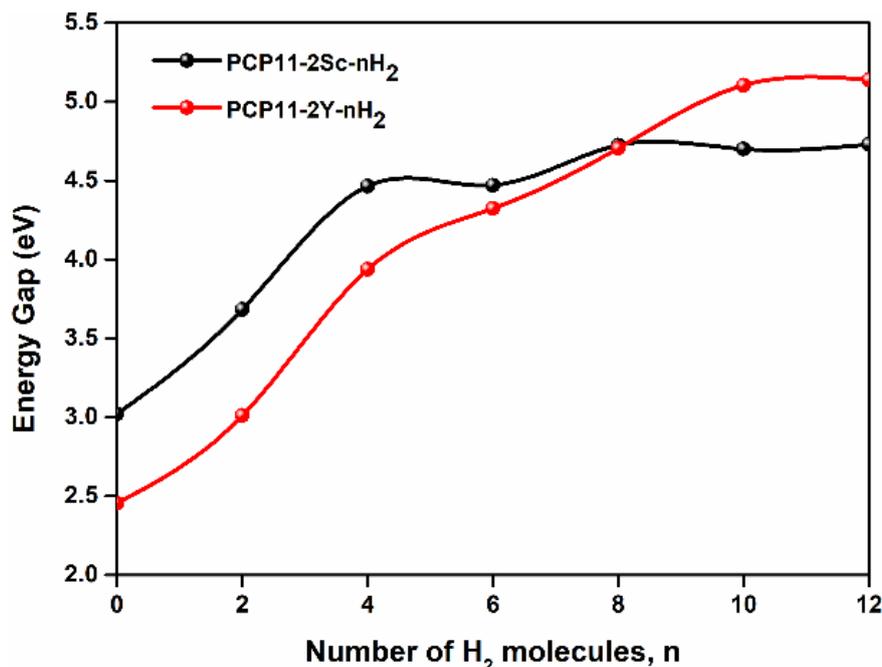

**Figure 4:** Variation of HOMO-LUMO gap with number of hydrogen molecules adsorbed on Sc/Y functionalized PCP11

### 3.3. Energetic parameters

To understand the hydrogen molecules interaction with TM atoms decorated PCP11, the average $H_2$ adsorption energy and successive $H_2$ desorption energy of studied systems are calculated using Equations 2 and 3 respectively. Figure 5 shows that, the average adsorption energy decreases with increase in number of $H_2$ adsorption. For first $2H_2$ adsorption on PCP11-2Sc, the $E_{ads}$ is found to be 0.39 eV, which decreases to 0.29 eV after attaining $H_2$ saturation ($12H_2$). This is attributed to the steric hindrance caused by $H_2$ crowed. The adsorption process involved in our study indicates that, the hydrogen molecules are adsorbed via Kubas-type of interaction with Sc atoms in which there is small charge donation from σ (HOMO) orbitals of hydrogen molecules to vacant 3d orbital of scandium atom occurs followed by back donation of charge from partial filled d orbitals of scandium to unfilled σ * (LUMO) orbitals of hydrogen molecules takes place [22, 62]. In this process, $H_2$ molecules gain a fraction of charge resulting in elongation of H-H bond lengths. The bonding of the last two $H_2$ molecules on Sc atoms is due to the charge polarization mechanism proposed by Niu-Rao-Jena [63, 64]. In this mechanism the positively charged Sc ion produces electric field which polarize the $H_2$ molecules and binds them in quasi-molecular manner. The average energy of hydrogen adsorption (0.36 eV) appears to be lower than the chemisorption but higher than the physisorption process which is a desired requirement for effective hydrogen storage as proposed by US-DOE. Similarly, For Y functionalized PCP11, the $1^{st}$ $2H_2$ molecules adsorbed with an average $E_{ads}$ of 0.35 eV, subsequently, the $2^{nd}$, $3^{rd}$, $4^{th}$, $5^{th}$ and $6^{th}$ $H_2$ adsorbed with $E_{ads}$ of 0.35, 0.34, 0.34, 0.33, and 0.28 eV respectively. The Y atoms also bind the $H_2$ molecules through Kubas-type interaction. Successive desorption energy (Figure: 5) decreases with increase in $H_2$ number which implies the fact that the outermost $H_2$ molecules require less energy to desorb than H$_2$ molecules close to TM centres and this is obvious due to weak polarization bonds.

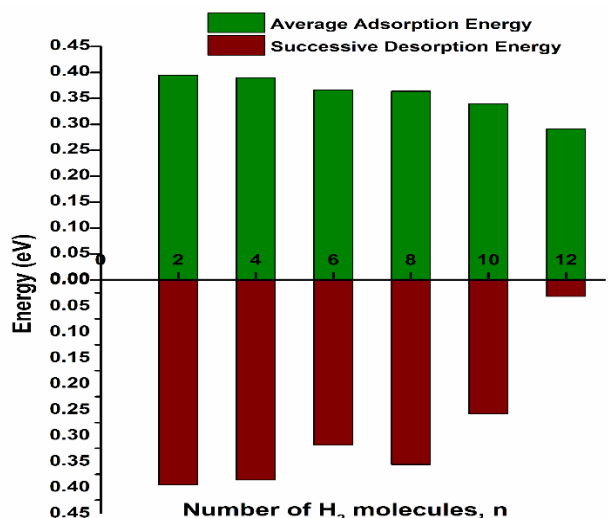

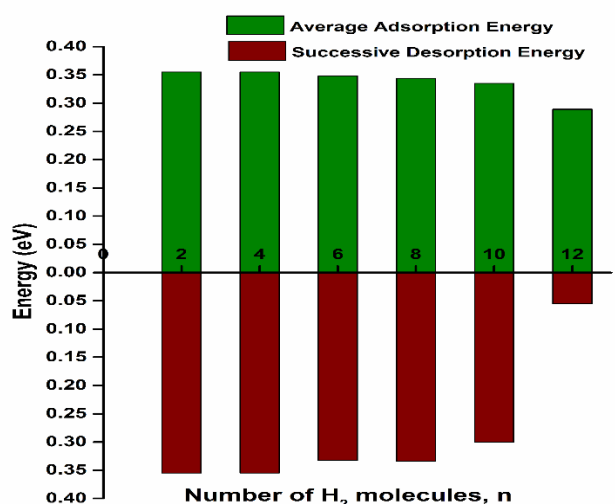

**Figure 5:** Average adsorption and successive desorption energy of (a) PCP11-2Sc-nH$_2$ (b) PCP11-2Y-nH$_2$

### 3.4. Electrostatic potential and Hirshfeld charges

To get an insight into the distribution of electron density over the system during the process of metal functionalization and hydrogen adsorption, electrostatic potential map on the total electron density have been plotted and depicted in Figure 6 and 7. The distribution of electron density is used to determine the active adsorption sites of the Sc and Y functionalized PCP11. In the ESP plot the region with red colour represents the accumulation of electron density (highest negative potential zone) whereas blue region represents the depletion of electron density (highest positive potential zone). The colour variation in ESP maps implies the transfer of electronic charge density during the adsorption of hydrogen [65]. The red region over the benzene ring represents the accumulation of electron density and with the functionalization of Sc and Y atoms over the benzene ring, the red regions disappear and a blue region appears over the metal atoms which indicate the depletion of electron density implying that the metal atom become somewhat ionic. It is observed from the ESP map of PCP11-2Sc and PCP11-2Y that the sorption centres (Sc and Y) are marked by dark bules surfaces indicating their electron deficiency as compared to PCP11. Upon addition of $H_2$ molecules to each sorption atoms, the colour map turns from dark blue to light blues implying the fact that the positive charges get transferred from the TM atoms to $H_2$ molecules. Successive addition of $H_2$ molecules to PCP11- 2Sc(2Y) changes the colour of TM atoms from blue to light blue and then to blueish-green indicating further transfer of charge density. From the figures, it is evident that the hydrogen saturated systems, PCP11-2Y(2Sc)-12$H_2$, have low accumulation of electron density over the top of $H_2$ and mild density of electron near benzene ring. This implies that more hydrogen molecules are unlikely to be adsorbed on the host material.

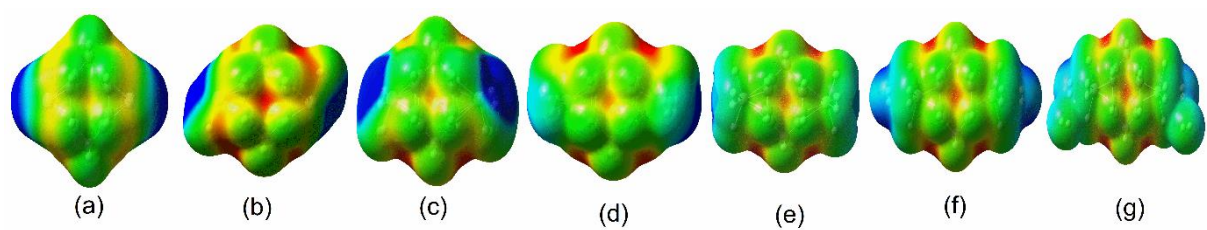

**Figure 6:** Electrostatics potential map of (a) PCP11-2Sc (b) PCP11-2Sc-2$H_2$ (c) PCP11-2Sc-4$H_2$ (d) PCP11-2Sc-6$H_2$ (e) PCP11-2Sc-8$H_2$ (f) PCP11-2Sc-10$H_2$ (g) PCP11-2Sc-12$H_2$

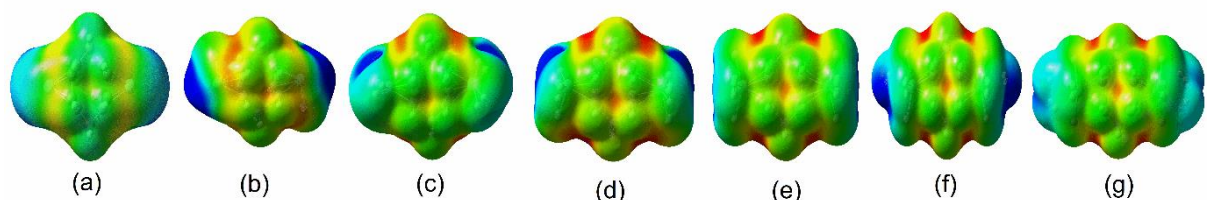

**Figure 7:** Electrostatics potential map of (a) PCP11-2Y (b) PCP11-2Y-2$H_2$ (c) PCP11-2Y-4$H_2$ (d) PCP11-2Y-6$H_2$ (e) PCP11-2Y-8$H_2$ (f) PCP11-2Y-10$H_2$ (g) PCP11-2Y-12$H_2$

To get a quantitative picture of the charge distribution of SC/Y functionalized PCP11 and $H_2$ stored system, the hirshfeld charge analysis has been performed. Figure 8 depicts the average hirshfeld charges on adsorbed $H_2$ molecules, transitions metal atoms (Sc, Y) and benzene ring carbon atoms as a function of adsorbed number of $H_2$ molecules. The hirshfeld charge on benzene ring C atoms of PCP11 is found to be -0.034 e which increases to -0.097 e after Sc functionalization. The charges on Sc in PCP11-2Sc(Y) is +0.42 e (+0.399 e) which infers that Sc (Y) atoms get more ionic during the PCP11 functionalization making them suitable adsorption centres. The average electronic charges on the benzene ring of PCP11-2Sc-2$H_2$ and PCP11-2Y-2$H_2$ is found to be 0.092 e and 0.095 e respectively, which reduces by 26% (for Sc) and 27% for hydrogen saturated Y functionalized PCP11. With successive adsorption of $H_2$ molecules on the systems, the average electronic charges on Sc and Y atoms increases by 38 % and 20% respectively. The average charges on adsorbed hydrogen molecules found in the range of -0.019 e to 0 e, this net charge gain by the $H_2$ molecules subject to a small elongation of H-H bond length.

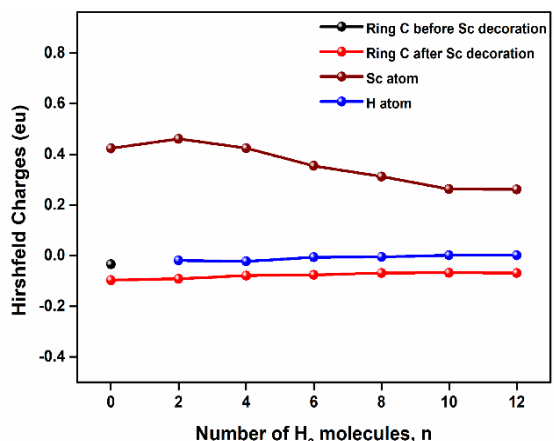

**(a)**

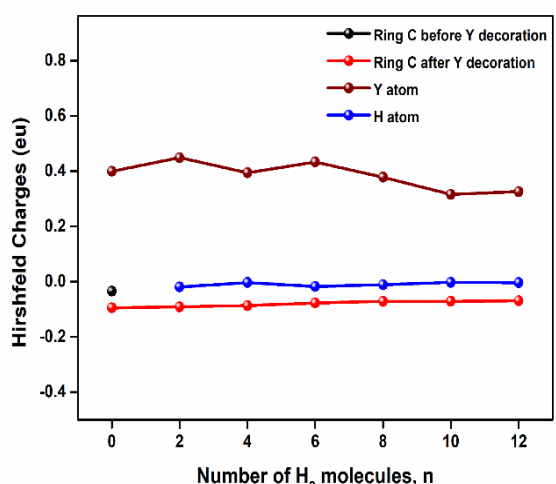

**(b)**

**Figure 8:** Hirshfeld charges before and after hydrogen adsorption on (a) PCP11-2Sc, (b)PCP11-2Y and adsorbed hydrogen

### 3.5. ADMP molecular dynamics simulations

To verify the hydrogen adsorption and desorption predicted by the DFT calculation, we have performed atomistic density matrix propagation (ADMP) simulations. ADMP belongs to the extended Lagrangian approach to molecular dynamics using gaussian basis functions and propagating the density matrix. Our calculation was performed with a canonical (NVT) ensemble by adding thermostat into ADMP. The nuclear kinetic energy thermostat is introduced by velocity scaling method and the temperature is checked and scaled at every 10-fs time step. For ADMP simulations, the maximum gravimetric density hydrogenated systems are kept at a pressure of 1 atm and three different temperatures above room temperature viz, 300 K, 375 K, and 475 K, for 1ps. The trajectory of potential energy and time evolution of bond length between sorption center and $H_2$ molecules are shown in Figure 9 and 10 respectively. ADMP simulations at 300 K illustrate that each Sc atom in PCP11-2Sc system, hold at least 4H$_2$ molecules just above room temperature, and the remaining 2H$_2$ desorbed from the host with the progress of time (Figure 10(a)). Similarly, in PCP11-2Y system, the last H$_2$ molecules get desorbed and at least 5H$_2$ molecules retains with each Y atoms at 300 K (Figure 10(b)). The adsorbed hydrogen molecules are oscillating near the sorption center within a distance of 1.9-2.5 Å for Sc and 2.1-3.3 Å Y functionalized system. As the temperature raised to 375 K and then to all hydrogen molecules from the vicinity of TM begins to increase movement after 500 fs (for Sc) and 350 fs (for Y). Since studied TM interacts strongly with hydrogen, the system needs higher temperature to desorb all H$_2$ molecules. At a temperature of 475 K, overall H$_2$ molecules began to move away from the sorption centers. The ADMP snapshots of both the studied system are shown in Figure *S5*.

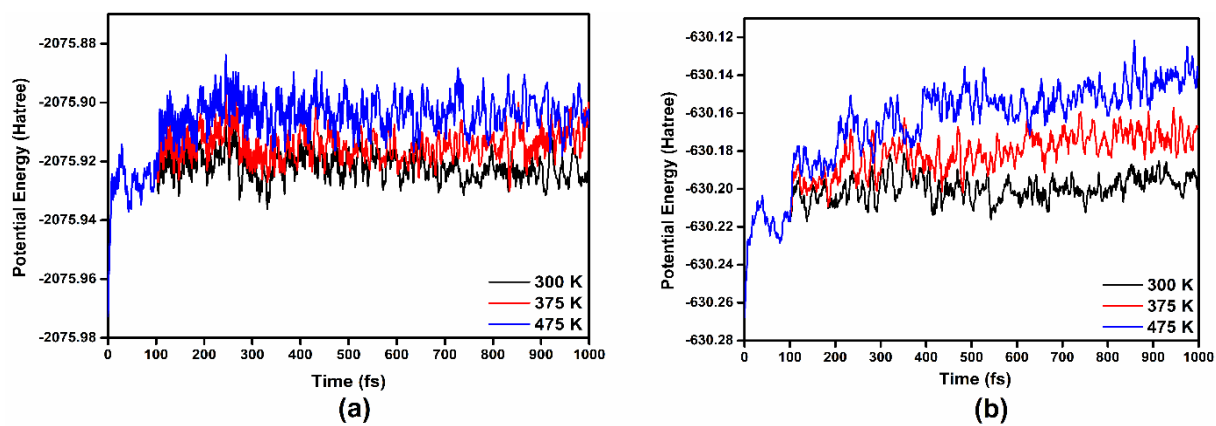

**Figure 9:** Potential energy trajectories of (a) Sc and (b) Y decorated PCP11 at 300K, 375K, and 475 K temperatures.

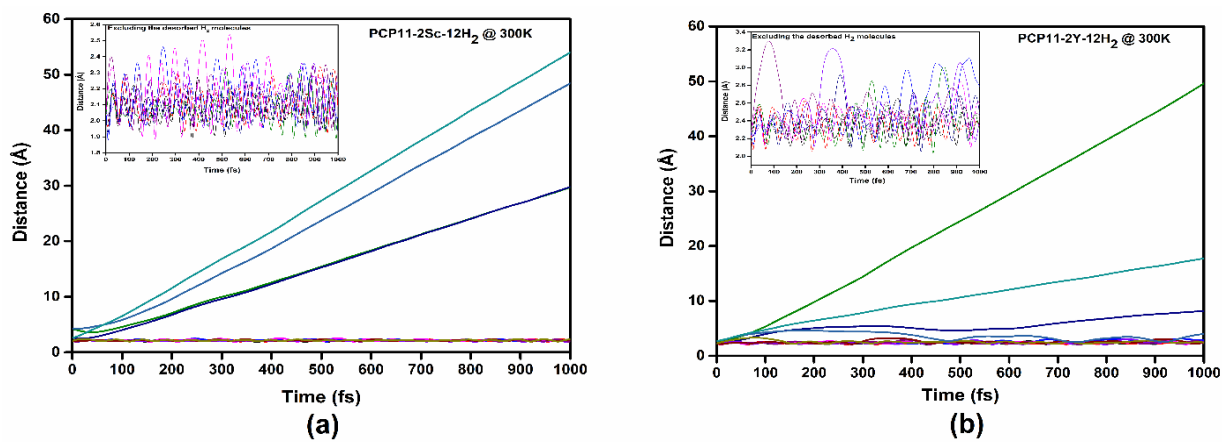

**Figure 10:** Time evolution trajectories of distance between TM and H$_2$ in (a) Sc and (b) Y functionalised PCP11 system at 300 K. (Each coloured line represents the distance of each adsorbed hydrogen molecules from sorption center)

For practical hydrogen storage substrate, we have examined the stabilities of Sc and Y decorated PCP11, at much above the room temperature (500 K) by performing the atomistic molecular dynamics simulations. It is recommended that for the realistic hydrogen storage, the H$_2$ molecules should be adsorbed on the substrates at room temperature and desorbed at a higher temperature. Secondly, the geometry of functionalized PCP11 should remain almost intact during the adsorption-desorption process. The atomistic molecular dynamics simulations for the above calculation have been performed for 1 ps with time step of 1 fs at a temperature of 500 K and 1 atm pressure. We observe that the Sc and Y decorated PCP11 structures are remain stable in the varied thermodynamic conditions, and negligible changes in C-C, C-Sc, and C-Y bond length after simulations are noted. The time evolution of C-Sc and C-Y bond lengths along with their potential energy trajectory are depicted in Figure S4 and S3 respectively. It is noticed that, the C-Sc and C-Y bond length oscillates around the mean value (for Sc: 2.20 Å and for Y: 2.42 Å) with negligible shift indicating the structural integrity above room temperature. Since these simulations ensure the solidity of host material, we believe that, studied materials can be proposed as feasible hydrogen storage system.

To avoid the metal-metal clustering, the diffusion energy barrier should be significantly greater than the thermal energy of transition metal (TM) atom (Sc and Y) at the highest desorption temperature. The thermal energy gained by the Sc and Y atoms at highest desorption temperature (500 K) is calculated using the following equation [1].

$$E = \frac{3}{2} K_B T$$

Here, E is the thermal energy of TM atoms, $K_B$ is the Boltzmann constant and T is the temperature of 500 K (higher than desorption temperature). We have calculated diffusion energy barrier for TM atoms by displacing the TM atoms from its stable position to next position. The calculated diffusion energy barrier (2.04 eV for Sc, and 1.83 For Y) is much higher than the obtained thermal energy of TM atoms (0.065 eV) at 500 K. These results prevent the possibility of metal clustering in the studied system.

### 3.6. Practical hydrogen capacity and gravimetric density

For a realistic hydrogen storage model, it is essential to estimate the adsorption and desorption of $H_2$ at a wide range of temperature (T) and pressure (P). Thus, it is more practical to calculate the number of hydrogen molecules available to use in an adsorption -desorption cycle. To get a quantitative picture of hydrogen storage (adsorption) and delivery (desorption) at different T and P, we calculated the number of hydrogen molecules adsorbed to each metal atom (occupation number) ( using equation 5 and 6) with the help of the empirical value of $H_2$ gas chemical potential ( μ ). Figure 11 depicts the plot of occupation number (*N*) for Sc/Y functionalized PCP11, using various value of μ at a range of temperature and pressure.

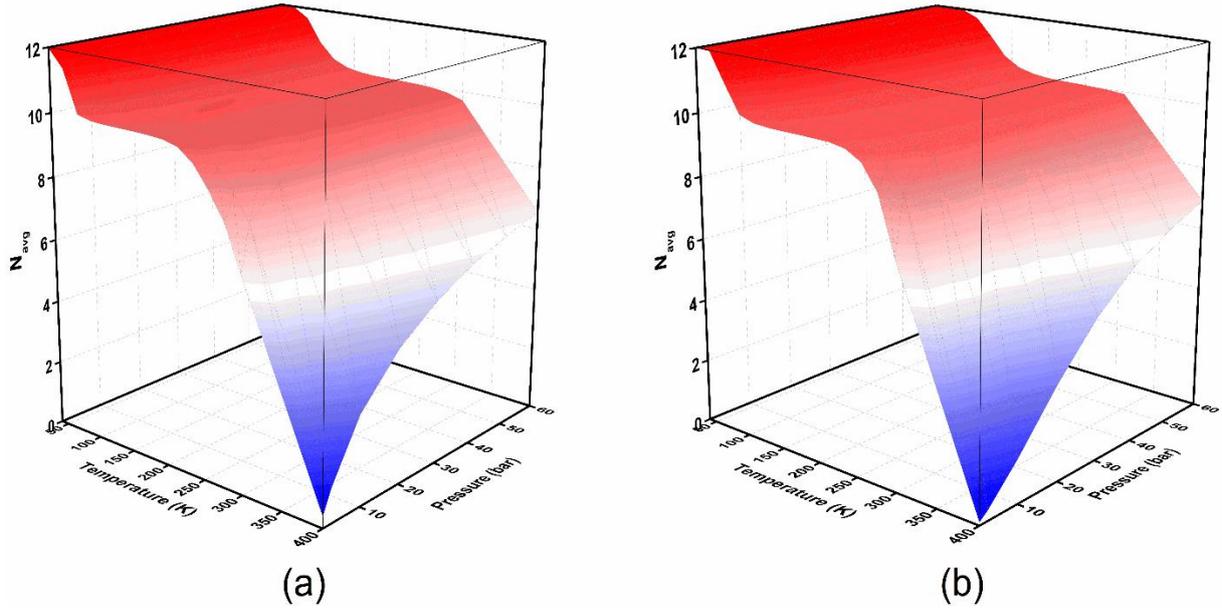

**Figure 11:** Hydrogen occupation number for (a) PCP11-2Sc and (b) PCP11-2Y at various T and P

Figure 11 shows that each Sc and Y atoms on PCP11 can hold 6$H_2$ molecules under temperature of 100K and a pressure range of 0-60 bar. At this condition the maximum gravimetric densities of Sc and Y decorated PCP11 system are calculated as 8.22 wt % and 6.33 wt % which is comparable with the experimentally reported value for Pd [38]. As the temperature increases beyond 100 K, $H_2$ molecules begin to desorb, while each Sc and Y atom has a maximum capacity to hold 5$H_2$ molecules throughout a temperature range of 100-300 K and pressure range of 0-60 bar. At a temperature of 400K under pressure range of 1-3 bar, almost all $H_2$ molecules from Y atoms are desorbed and PCP11-2Sc finally carries ~1$H_2$. while plotting the of occupation number, we considered two thermodynamic conditions viz: ideal storage condition (300 K under 30 bar) and ideal delivery condition (400K under 3 bar). For PCP11-2Sc system, at 300 K and 30 bar, *N* is calculated to be 9.50. This N value signifies that, each Sc atom carries around 5$H_2$ molecules resulting in the gravimetric density of 6.62 wt %. At 400 K under 3 bar, *N* is found to be 1.06which implies that only one $H_2$ molecule is retained by the host at this condition. For PCP11-2Y system, each Y atom carries 5$H_2$ molecules at storage condition giving rise to 5.26 wt% and desorb all $H_2$ molecules at the delivery condition. The number of usable hydrogen molecules for Sc and Y functionalized PCP11 are thus found to be around 8$H_2$ and 10$H_2$ respectively giving rise to the gravimetric densities higher than the US-DOE target. This makes us propose our materials suitable for practical hydrogen storage.

**Table 3:** Results of thermodynamically usable hydrogen storage capacity

| System | $N_{theory}$ | $N_{ads}$ | $N_{des}$ | $N_{use}$ | $G_{theory}$ | $G_{practical}$ | $G_{usable}$ |
|---|---|---|---|---|---|---|---|
| PCP11-2Sc-2$H_2$ | 2 | 1.99 | 0.70 | 1.29 | 1.47 | 1.46 | 0.95 |
| PCP11-2Sc-4$H_2$ | 4 | 3.99 | 1.02 | 2.97 | 2.9 | 2.89 | 2.17 |
| PCP11-2Sc-6$H_2$ | 6 | 5.96 | 1.06 | 4.9 | 4.28 | 4.26 | 3.52 |
| PCP11-2Sc-8H$_2$ | 8 | 7.95 | 1.06 | 6.89 | 5.63 | 5.60 | 4.89 |
| PCP11-2Sc-10H$_2$ | 10 | 9.48 | 1.06 | 8.42 | 6.94 | 6.61 | 5.91 |
| PCP11-2Sc-12H$_2$ | 12 | 9.50 | 1.06 | 8.44 | 8.22 | 6.62 | 5.92 |
| PCP11-2Y-2H$_2$ | 2 | 1.98 | 0.26 | 1.72 | 1.11 | 1.10 | 0.96 |
| PCP11-2Y-4H$_2$ | 4 | 3.98 | 0.33 | 3.65 | 2.20 | 2.19 | 2.01 |
| PCP11-2Y-6H$_2$ | 6 | 5.96 | 0.33 | 5.63 | 3.27 | 3.25 | 3.07 |
| PCP11-2Y-8H$_2$ | 8 | 7.96 | 0.34 | 7.62 | 4.31 | 4.29 | 4.11 |
| PCP11-2Y-10H$_2$ | 10 | 9.86 | 0.34 | 9.52 | 5.33 | 5.26 | 5.09 |
| PCP11-2Y-12H$_2$ | 12 | 9.86 | 0.34 | 9.52 | 6.33 | 5.26 | 5.09 |

*[ $N_{theory}$ is the number of theoretically adsorbed H$_2$ molecules in DFT. $N_{ads}$ and $N_{des}$ are the number of H$_2$ molecules adsorbed at storage (300 K - 60 bar) and delivery (400 K - 3 bar) conditions respectively. $N_{use}$ is the difference between $N_{ads}$ and $N_{des}$, represent the usable number of H$_2$ molecules. $G_{theory}$ and $G_{practical}$ are theoretical and practical hydrogen wt% at storage-delivery conditions respectively. $G_{usable}$ is the wt% of hydrogen that can be use practically.]*

Now, while considering the hydrogen storage systems for practical use such as vehicular purposes, characterization of the desorption temperature plays an important role. We calculated the prevailing desorption temperatures of the studied systems by using Van't Hoff equation as described below,

$$T_D = \left(\frac{E_{ads}}{K_B}\right)\left(\frac{\Delta S}{R} - \ln p\right)^{-1} \qquad (7)$$

Here $E_{ads}$ is the calculated hydrogen adsorption energy with Sc and Y decorated PCP11. $K_B$ is the Boltzmann constant, R is the gas constant, P is the equilibrium pressure (in our calculation we use 1 to 5 atm with increment of 0.5 atm) and $\Delta S$ is the change in entropy of hydrogen from gas to liquid phase [66]. The maximum desorption temperature ($T_{D[max]}$) is calculated by using the adsorption energy of first 2H$_2$ molecules adsorbed by the host materials. This is essential to understand the complete delivery of hydrogen from the studied system. By using the adsorption energy of sixth H$_2$ molecules adsorbed by each Sc and Y atom on PCP11, the minimum desorption temperature ($T_{D[min]}$) is calculated, which is necessary to activate the desorption process of H$_2$ from the host. Figure 12 shows the Van't Hoff desorption temperature with a range of equilibrium pressure of 1-5 atm. At a pressure of 1 atm the calculated minimum desorption temperature for Sc and Y doped PCP11 are 372 K and 370 K respectively. The desorption temperature for hydrogenated systems are increases with the increase in equilibrium pressure. The estimated value of average desorption temperature confirms that, all the adsorbed H$_2$ molecules are not dissociate at room temperature and even in small thermal fluctuation.

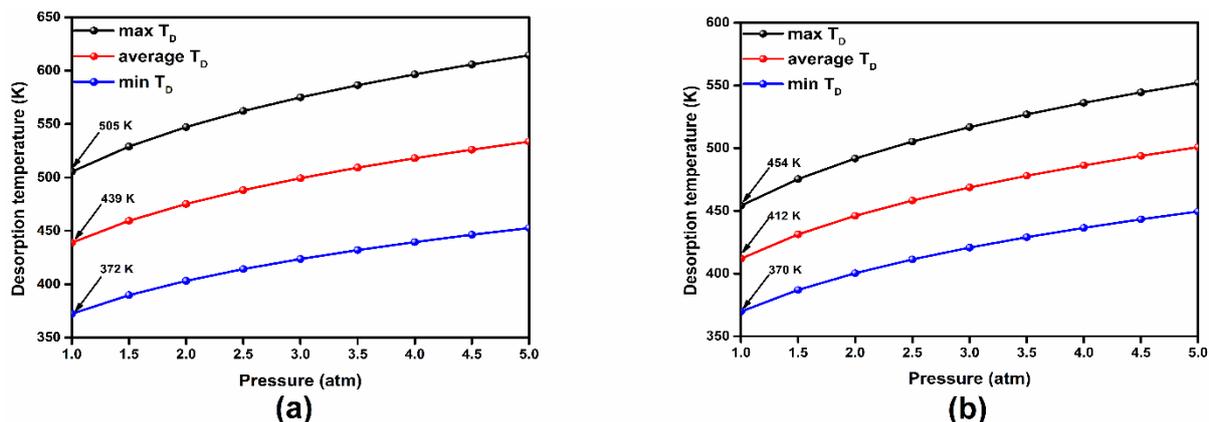

**Figure 12:** Plot of Vant-Hoff desorption temperature for (a) PCP11-2Sc-nH$_2$ and (b) PCP11-2Y-nH$_2$

## 4. CONCLUSION

We investigated the hydrogen storage capacity of scandium and yttrium functionalized [1,1]paracyclophane using density functional theory calculations. Transition metal atoms (Scandium and yttrium) are found to strongly bind with PCP11 through Dewar mechanism. Each of Sc and Y atoms on PCP11 were found to adsorb up to 6H$_2$ molecules via Kubas-type interaction with an average adsorption energy of 0.36 eV and 0.34 eV respectively which was intermediate between chemisorption and physisorption. Average desorption temperature for Sc and Y decorated system was calculated to be 439 K and 412 K respectively at 1 atm of pressure. For Sc and Y functionalized PCP11, the calculated hydrogen storage capacities are 8.22 wt% and 6.33 wt% respectively, however at the adsorption thermodynamics of 300 K and 30 bar the 6.62 wt% and 5.26 wt% respectively. At desorption condition of 400 K and 3 bar, the practical usable hydrogen became 5.92% and 5.09 wt% which is good enough as per US-DOE target. ADMP molecular dynamics simulations showed that the desorption of hydrogen was sufficiently above room temperature (500K) which is quite essential for practical use of hydrogen storage. Scandium and yttrium functionalized PCP11 have suitable adsorption energy and desorption temperature for H$_2$ molecules, the host structure is stable above desorption temperature and good hydrogen storage capacity. Therefore, the Sc and Y functionalized PCP11 can be proposed as practically viable systems for reversible hydrogen storage at ambient environment. We hope that our work can contribute substantially to the research and development of new materials for hydrogen energy and help experimental synthesis of the material.

## 5. ACKNOWLEDGMENT

Authors acknowledges Indian Institute of Technology (Indian School of Mines), Dhanbad for providing support and other research facilities. We acknowledge National Supercomputing Mission (NSM) for providing partial computing resources of 'PARAM Shivay' at Indian Institute of Technology (BHU), Varanasi.